%% file: QCD-bei4.tex
\documentclass{ws-ijmpa}
\begin{document}
\newcommand{\be}{\begin{eqnarray}}
\newcommand{\ee}{\end{eqnarray}}
\newcommand\del{\partial}
\newcommand{\nn}{\nonumber } 
\markboth{J.C. Osborn}{Chiral Symmetry Breaking}

\catchline{}{}{}{}{}

\title{CHIRAL SYMMETRY BREAKING AT NONZERO CHEMICAL POTENTIAL}

\author{J.C. Osborn}
\address
{Physics Department, Boston University,
Boston, MA 02215, USA}

\author{K. Splittorff}
\address
{Niels Bohr Institute, Blegdamsvej 17, DK-2100, Copenhagen {\O}, Denmark}

\author{J.J.M. Verbaarschot}
\address{University at Stony Brook, Stony Brook, NY 11794, USA.}  


\maketitle


\begin{abstract}
We consider chiral symmetry breaking at nonzero chemical potential and
discuss the relation with the spectrum of the Dirac operator.
We solve the so called Silver Blaze Problem that the chiral condensate
at zero temperature does not depend on the chemical potential
while this is not the case for 
the Dirac spectrum and the weight of the partition function.

\end{abstract}

\keywords{Chiral Symmetry, Dirac Spectrum, Nonzero Chemical Potential}


\section{Introduction}

The QCD phase diagram has been explored in great detail during
the past decade. New phases have been discovered and a variety of parameters
have been considered \cite{aw,ss,sss,KT,thomas,krishna,kogut-misha}. 
In particular, a great deal of progress has been made for QCD
at large nonzero baryon chemical potential (denoted by $\mu$). 
Based on model calculations
and asymptotic expansions, it is now generally accepted
that the ground state of QCD in this region is in a phase with a
diquark condensate in a color-flavor locked
state\cite{krishna2}. 
The QCD phase diagram is best known at $\mu=0$
where it has been explored extensively by
lattice simulations\cite{lattmu}. At a critical 
temperature of about 160 MeV we expect a cross-over from a phase
with a nonzero chiral condensate to phase with restored chiral symmetry.
At zero temperature, the
QCD partition function is independent of  $\mu$
until it has reached the lightest excitation
with nonzero baryon number. This means that the chiral condensate
remains constant up to $\mu = m_N/3-\epsilon$ 
(with $\epsilon$ the nuclear binding energy). 
The transition to
the chirally restored phase is believed to be of first order at
low temperatures and ends in a
critical endpoint\cite{Barda,berges,adam,Luo}.
A schematic QCD phase diagram is given in Fig. 1.
 \begin{figure}[!t]
         \centerline{
           \scalebox{0.3}{
             \input{phidia.pstex_t}  }}
         \caption{Schematic phase diagram 
of the QCD partition function at nonzero
temperature and chemical potential. The short curve denotes 
the nuclear matter liquid gas transition.}
         \label{fig-eg}
       \end{figure}
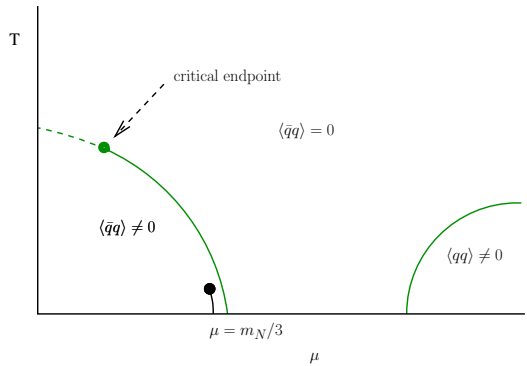
In this lecture we will consider the QCD partition function at zero
temperature for chemical potentials $\mu \ll m_N/3$, where QCD can be
described  by a chiral Lagrangian.

\section{The QCD Partition Function}

\begin{figure}[!t]
         \centerline{
           \scalebox{0.3}{
             \input{zmu.pstex_t} }}
         \caption{The QCD partition function as a function of
the chemical potential for zero and  nonzero
temperature in the absence of nuclear binding energy. 
The temperature of the critical endpoint in Fig. 1
is denoted by $T_{\rm end}$. }
         \label{fig2}
       \end{figure}
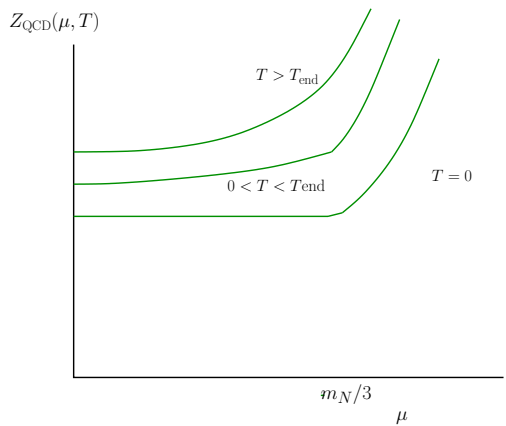

The QCD partition function at temperature $1/\beta$ is given by
\be
Z_{\rm QCD}(\mu,T) = \sum_k e^{-\beta(E_k -\mu N_k)},
\ee
where the sum is over all quantum states of QCD with energy $E_k$ and
quark number  $N_k$.
The partition function is a
monotonously nondecreasing function of the chemical
potential. For $T=0$ its value remains constant up to 
$\mu = m_N/3 - \epsilon$
(see Fig. \ref{fig2}). This implies that thermodynamic observables, such
as for example the chiral condensate,
do not depend on the chemical potential up to this point.
However, this simple fact
becomes a mystery when we consider the 
Euclidean QCD partition function given by (with quark masses denoted by $m_f$)
\be
Z_{\rm QCD}(\mu,T) 
= \langle \prod_{f=1}^{N_f} \det (D +\mu \gamma_0 +m_f) \rangle
= \langle \prod_{f=1}^{N_f} 
\prod_k (\lambda_k +m_f)\rangle.
\ee
Here and below, the brackets denote the
average over gauge field configurations weighted by the
Euclidean Yang-Mills action. The quark determinant has
been expressed in terms of the eigenvalues of the Dirac operator
given by
\be
(D+\mu \gamma_0)\phi_k = \lambda_k \phi_k.
\ee
The chiral condensate is thus given by (we only consider $m_f = m$ from
now on)
\be
|\langle \bar q q\rangle| = 
\lim \frac 1V \frac  1{N_f}\del_m \log Z_{\rm QCD}(\mu,T)
=\lim \frac 1V \left \langle \sum_k \frac 1{\lambda_k +m} 
\right\rangle_{\rm QCD}.
\label{condens}
\ee
The mystery is that the eigenvalues and the quark determinant
strongly depend on the chemical potential but
the partition function and the chiral condensate do not. 
Thomas Cohen\cite{cohen} coined this problem as
the Silver Blaze problem after the title of a novel of Conrad Doyle
where a racing horse with the name of the title disappeared without
the dogs barking. The aim of the lecture is
to elucidate this problem.

\section{Solution of the Silver Blaze Problem}

The spectral ``density'' of the Dirac operator is given by
\be
\rho(z,z^*,m,\mu) = \langle \sum_k \delta^2(z-\lambda_k)
\det(D+\mu\gamma_0+m)^{N_f}\rangle.
\ee
The quenched spectral density is obtained for $N_f=0$.
We also use the resolvent defined by (here and below we
take the volume $V$ finite but large)
\be
G(z,z^*,m,\mu) = \langle \frac 1V \sum_k \frac 1{z+\lambda_k}
\det(D+\mu\gamma_0+m)^{N_f}\rangle,
\ee
so that (notice that $\rho(-z,-z^*,m,\mu) =\rho(z,z^*,m,\mu)$)
\be
\frac{\rho(z,z^*, m,\mu)}V =\frac 1{\pi} \del_{z^*} G(z,z^*,m,\mu).
\label{constraintg}
\ee
Since the chiral condensate is given by,
\be
|\langle \bar q q\rangle| &=& \int d^2 z \frac 1V 
\frac 1{z+m} \rho(z,z^*,m,\mu) = G(z=m,z^*=m, m,\mu)
\label{condens2}
\ee
we have solved the Silver Blaze problem if we can show that the 
spectral density satisfies the constraint (\ref{constraintg}) with 
$G(z=m,z^*=m, m,\mu)$ independent of $\mu$. 
 At $\mu=0$, 
the discontinuity of the chiral condensate is due
to an accumulation of eigenvalues on the imaginary axis 
(Banks-Casher relation). 
At $\mu \ne 0$ the QCD Dirac operator
does not have any hermiticity properties. The eigenvalues are scattered
in the complex plane and, because of the phase of the fermion determinant,
the eigenvalue density is complex. 
The solution of the Silver
Blaze problem will provide us with different 
relation between the discontinuity
of the chiral condensate and the spectral density.

\section{The Dirac Spectrum at Nonzero Chemical Potential}

The spectrum of the quenched Dirac operator is
well understood\cite{everybody,misha,janik,HOV,TV,tilo0,tilo}.  
In the mean field limit
limit at fixed $\mu \ll \Lambda_{\rm QCD}$ the eigenvalues are scattered
uniformly in the strip $ |{\rm Re}(z)| < 2\mu^2F^2/\Sigma $. 
The mean field limit of the quenched resolvent
is therefore given by 
\be
G_{\rm Q}(z,z^*,\mu) &=& \frac { {\rm Re} (z)\Sigma^2} {2\mu^2 F^2}\qquad {\rm for } \qquad 
|{\rm Re}(z)| <2\mu^2F^2/\Sigma, \label{pig}\\
G_{\rm Q}(z,z^*,\mu) &=& \Sigma \qquad {\rm for } \qquad 
|{\rm Re}(z)|>2\mu^2F^2/\Sigma.
\ee
In the unquenched case we do not have a simple expression
for the resolvent except that it has to satisfy the
constraint (\ref{constraintg}) at zero temperature.

The spectral density of the Dirac operator can be  obtained 
from\cite{misha,Feinh} 
\be
\rho(z,z^*,m,\mu) = \lim_{n\to 0}\frac 1{\pi n} \del_z\del_{z^*} 
\log Z_n(z,z^*,m,\mu)\ee
with generating function $Z_n(z,z^*,\mu)$ defined by\cite{misha,TV}
\be
Z_n(z,z^*,m,\mu) &=& \langle \det(D+\mu\gamma_0 +m) {\det}^n(D+\mu\gamma_0 +z) 
{\det}^n(D^\dagger+\mu\gamma_0 +z^*)\rangle \nn\\
&=& \langle \det(D+\mu\gamma_0 +m) {\det}^n(D+\mu\gamma_0 +z) 
{\det}^n(D-\mu\gamma_0 +z^*)\rangle. \qquad
\label{genfun}
\ee 
In general, the replica limit $n \to 0$ is not well defined\cite{critique}.
However, for $m, |z| \ll F^2/(\Sigma \sqrt V)$ and $\mu \ll \Lambda_{\rm QCD}$
this family of partition functions satisfies the
Toda lattice hierarchy which allows one to obtain the $n \to 0$ result
from a recursion relation\cite{kanzieper,splitv1} both in the 
quenched\cite{splitv2} and in the unquenched case\cite{AOSV}. The
unquenched spectral density, which 
was first obtained\cite{james} from an equivalent
random matrix model,
is a strongly oscillating complex function (see
Fig. 3). The amplitude of the oscillations grows
exponentially with $\mu^2F^2V$ while the period goes as
$1/\Sigma V$. 
It can be
shown\cite{OSV} that the resolvent, obtained by integrating the oscillating
spectral density,
satisfies the constraint that $G(z=m,z^*=m,m,\mu)$ is independent of
$\mu$. The discontinuity of the chiral
condensate is $\mu$-independent with  contributions from
the entire oscillating region\cite{OSV}. 
This solves the Silver Blaze problem.
\begin{figure}[!t]
\label{fig3}
\unitlength 1cm 
  \unitlength1.0cm
  \begin{center}
\begin{picture}(14.0, 6.0)
  \psfig{file=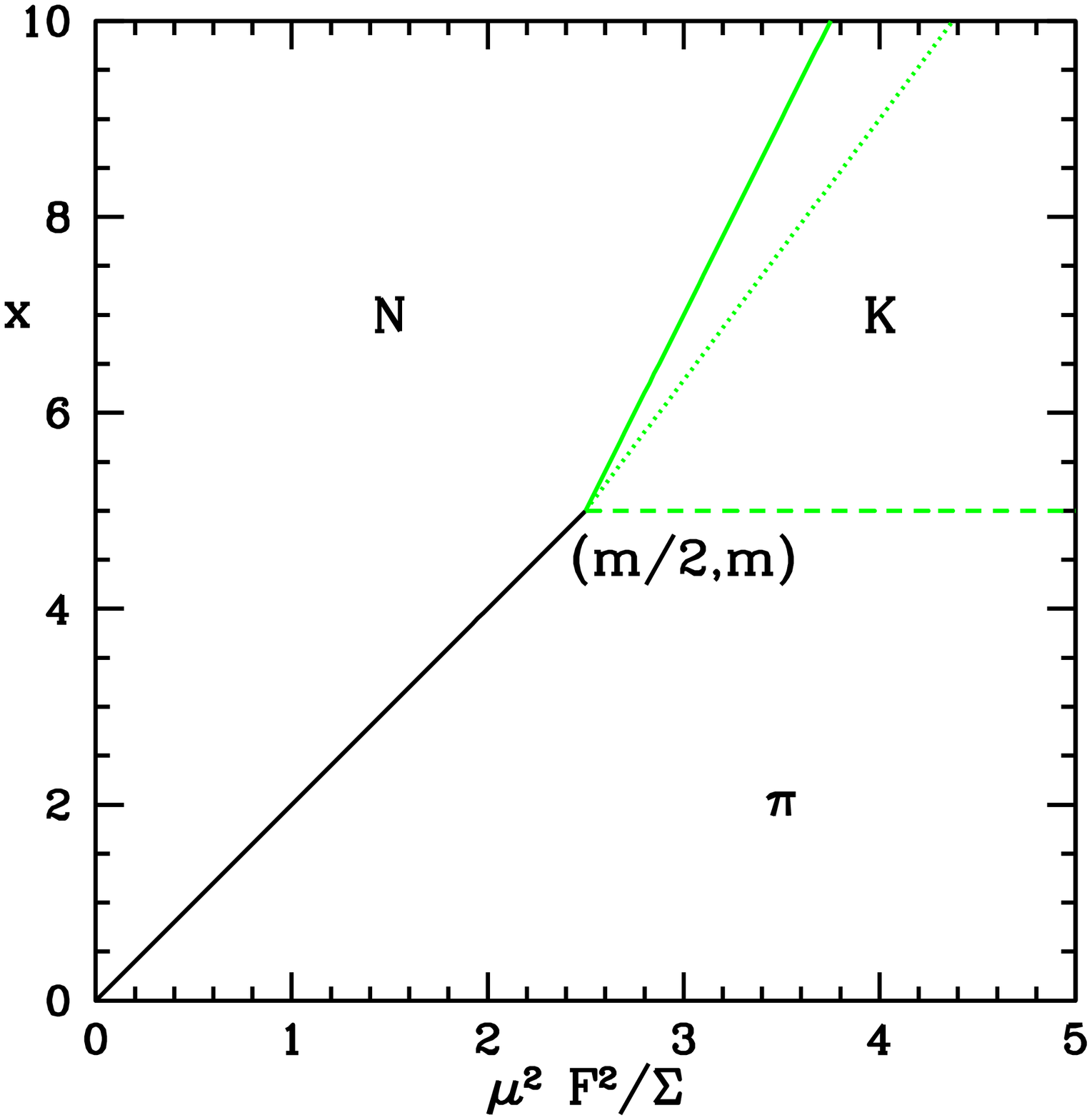,width=6.5cm}
 \psfig{file=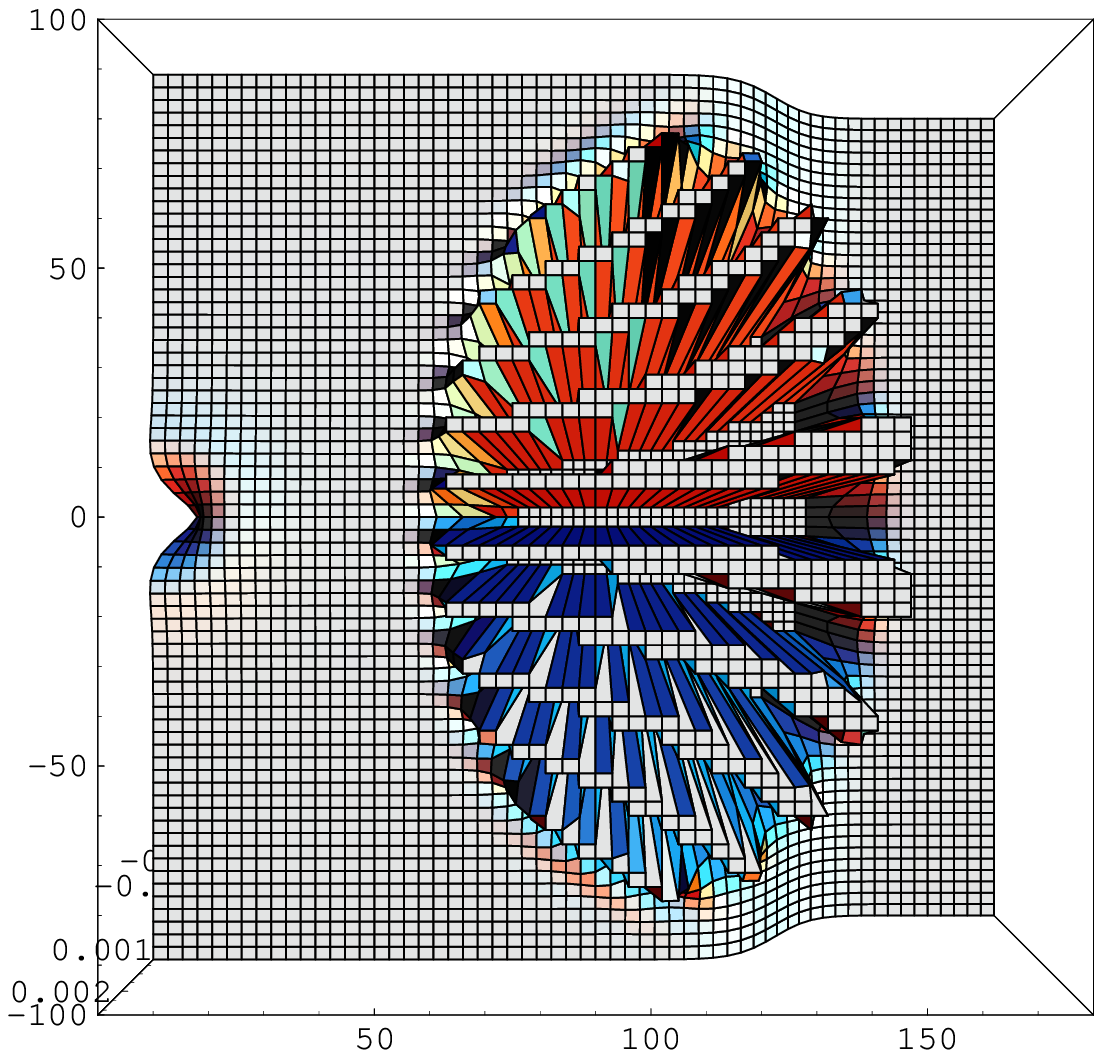,width=6.3cm}
\put(-2.0,-0.3){{\bf$ x\Sigma V$}}
\put(-6.6,4.0){{\bf $y\Sigma V$}} 
\put(-6.7,-0.3){{\bf ${\rm Re}[ \rho]/\Sigma^2 V^2$}}
\end{picture}
\end{center}
\caption[]{Phase diagram of Dirac spectrum (left) and 
real part of the spectral density for full QCD at
nonzero chemical potential (right\cite{OSV}). 
In the left figure full lines denote
a second order transition and dashed curves a first order transition. The
dotted curve ($x=\frac 83 \mu^2F^2/\Sigma - \frac m3$) 
represents the right edge of the oscillation region given in
the right figure.}
\end{figure}

The spectral density can also be evaluated in the
mean field limit. Then
the resolvent does not depend on the
number of replicas\cite{critique,misha,OSV2}
and the replica limit can be taken
trivially. 
Assuming that this procedure gives the correct result we
simply consider $n=1$. The mean field generating function
(\ref{genfun}) for $n=1$ and real $z$ was analyzed before\cite{KT} by means
of a chiral Lagrangian. Interpreting $m$ as the strange
quark mass and $z=z^*$ as the light quark masses, the valence ``pion'' 
and ``kaon'' masses are given by,
\be
m_{{\rm v} \, \pi}^2 = \frac {2x \Sigma}{F^2}\qquad {\rm and}
 \qquad m_{{\rm v} \, K}^2 = \frac {(x+m) \Sigma}{F^2} \qquad {\rm
   with} \qquad  x={\rm Re}(z)
\ee
respectively. With this identification, the isospin chemical
potential is equal to $\mu_I=2\mu$ and
the strangeness chemical potential $\mu_s=-\mu$.
The phase diagram in the $x-\mu^2$ plane   for Im$(z)=0$ is shown
in the left panel of Fig. 3.
We observe three different phases, a normal phase ($N$), a pion condensation
phase ($\pi$) and a kaon condensation phase ($K$) bordered by the
dashed curve and the solid curve $x=4\mu^2F^2/\Sigma-m$. At fixed chemical
potential, a transition from the $\pi$-phase to the $K$-phase takes
place when $m_{{\rm v} \, \pi}=m_{{\rm v} \, K}$.
Evaluating the mean field resolvent in each of the phases we find
\be
G_N = \Sigma, \quad G_K=\Sigma, \quad G_\pi = \frac{{\rm Re} (z)\Sigma^2}
{2\mu^2F^2}.
\ee
The resulting spectral density is zero in the
$N$-phase  
and a constant in the $\pi$-phase in agreement with the
thermodynamic limit of the exact spectral density. However, the mean
field spectral density in the $K$-phase is zero rather than
oscillatory.
The discontinuity of the mean field resolvent across the interface
of the $\pi$-phase and the $K$-phase gives a delta function in the
spectral density. As the quark mass approaches zero, the interface
approaches zero resulting in a discontinuity
of the chiral condensate due to an accumulation of eigenvalues.
This solves the Silver Blaze problem for the mean field spectral density for 
$N_f =1$ obtained from $n=1$.

\section{Discussion} 

We have described two different mechanisms to obtain a discontinuity
in the chiral condensate at nonzero chemical potential. 
First, in case of the low energy limit of
the full QCD partition function,
the discontinuity arises from the
oscillating part of the spectral density.
Second, using a replica mean field result for the spectral density
we find that the discontinuity moves with 
$m$ and reaches the imaginary axis for $m =0$. 
 At this moment it is not
yet clear whether or how these two pictures are related. 
One important observation
is that the oscillating region of the spectral density
is absent for $\mu < m_\pi/2$. We expect\cite{kim} that the sign problem 
is less severe this in this region of the phase diagram which could open
the door to  realistic lattice simulations\cite{fodor}.

\noindent{\bf Acknowledgments}. Gernot Akemann and Tilo Wettig are
thanked for useful discussions. Philip de Forcrand is acknowledged
for the suggestion to adopt the name ``The Silver Blaze Problem''
as suggested by Tom Cohen. This work was partially supported
by U.S. DOE grant DE-FC02-01ER41180 (JCO) 
and by U.S. DOE grant DE-FG-88ER40388 (JJMV).


\end{document}

%% file: phidia.pstex_t
\begin{picture}(0,0)%
\includegraphics{phidia.pstex}%
\end{picture}%
\setlength{\unitlength}{3947sp}%
\begingroup\makeatletter\ifx\SetFigFont\undefined%
\gdef\SetFigFont#1#2#3#4#5{%
  \reset@font\fontsize{#1}{#2pt}%
  \fontfamily{#3}\fontseries{#4}\fontshape{#5}%
  \selectfont}%
\fi\endgroup%
\begin{picture}(10833,8291)(826,-7580)
\put(4276,-1636){\makebox(0,0)[lb]{\smash{{\SetFigFont{20}{24.0}{\familydefault}{\mddefault}{\updefault}critical endpoint}}}}
\put(2701,-4786){\makebox(0,0)[lb]{\smash{{\SetFigFont{20}{24.0}{\familydefault}{\mddefault}{\updefault}$\langle \bar q q \rangle \ne 0$}}}}
\put(2701,-4786){\makebox(0,0)[lb]{\smash{{\SetFigFont{20}{24.0}{\familydefault}{\mddefault}{\updefault}$\langle \bar q q \rangle \ne 0$}}}}
\put(9976,-5386){\makebox(0,0)[lb]{\smash{{\SetFigFont{20}{24.0}{\familydefault}{\mddefault}{\updefault}$\langle q q \rangle \ne 0$}}}}
\put(7126,-7486){\makebox(0,0)[lb]{\smash{{\SetFigFont{20}{24.0}{\familydefault}{\mddefault}{\updefault}$\mu$}}}}
\put(6451,-2761){\makebox(0,0)[lb]{\smash{{\SetFigFont{20}{24.0}{\familydefault}{\mddefault}{\updefault}$\langle \bar q q \rangle = 0$}}}}
\put(5026,-6961){\makebox(0,0)[lb]{\smash{{\SetFigFont{20}{24.0}{\rmdefault}{\mddefault}{\updefault}$\mu =m_N/3$}}}}
\end{picture}%

%% file: zmu.pstex_t
\begin{picture}(0,0)%
\includegraphics{zmu.pstex}%
\end{picture}%
\setlength{\unitlength}{3947sp}%
\begingroup\makeatletter\ifx\SetFigFont\undefined%
\gdef\SetFigFont#1#2#3#4#5{%
  \reset@font\fontsize{#1}{#2pt}%
  \fontfamily{#3}\fontseries{#4}\fontshape{#5}%
  \selectfont}%
\fi\endgroup%
\begin{picture}(10383,8784)(301,-8212)
\put(6826,-7636){\makebox(0,0)[lb]{\smash{{\SetFigFont{29}{34.8}{\familydefault}{\mddefault}{\updefault}$m_N/3$}}}}
\put(8401,-8086){\makebox(0,0)[lb]{\smash{{\SetFigFont{29}{34.8}{\familydefault}{\mddefault}{\updefault}$\mu$}}}}
\put(9151,-3061){\makebox(0,0)[lb]{\smash{{\SetFigFont{20}{24.0}{\familydefault}{\mddefault}{\updefault}$T=0$}}}}
\put(4876,-3286){\makebox(0,0)[lb]{\smash{{\SetFigFont{20}{24.0}{\familydefault}{\mddefault}{\updefault}$0<T<T{\rm end}$}}}}
\put(5476,-961){\makebox(0,0)[lb]{\smash{{\SetFigFont{20}{24.0}{\familydefault}{\mddefault}{\updefault}$T>T_{\rm end}$}}}}
\put(301,164){\makebox(0,0)[lb]{\smash{{\SetFigFont{25}{30.0}{\rmdefault}{\mddefault}{\updefault}$Z_{\rm QCD}(\mu,T)$}}}}
\end{picture}%